\documentclass[journal]{IEEEtran}
\IEEEoverridecommandlockouts

\usepackage{xfrac}
\usepackage{amsmath}
\usepackage{multirow,etoolbox}
\usepackage{color}
\usepackage{amssymb}
\usepackage{cite}
\usepackage{amssymb,amsmath,mathtools}

\usepackage{diagbox}

\usepackage{subcaption}
\usepackage[utf8]{inputenc}

\usepackage[subtle]{savetrees}

\usepackage{stfloats}
\usepackage{relsize}

\newcommand{\MeijerG}[7]{G \begin{smallmatrix} #1 & #2 \\ #3 & #4 \end{smallmatrix} \left( \begin{smallmatrix} #5 \\ #6 \end{smallmatrix} \middle\vert #7 \right) }

\makeatletter
\patchcmd{\@maketitle}
  {\addvspace{0.5\baselineskip}\egroup}
  {\addvspace{-1\baselineskip}\egroup}
  {}
  {}
\makeatother

\begin{document}
\bstctlcite{IEEEexample:BSTcontrol}

\title{A Hybrid Energy Harvesting Protocol for Cooperative NOMA: Error Performance Approach}
\author{Faical Khennoufa, Khelil Abdellatif, Ferdi Kara,~\IEEEmembership{Senior Member,~IEEE,} Hakan Kaya, Xingwang Li, ~\IEEEmembership{Senior Member,~IEEE,} Khaled Rabie, ~\IEEEmembership{Senior Member,~IEEE,} and  Halim Yanikomeroglu, ~\IEEEmembership{Fellow,~IEEE.} 
\thanks{ F. Khennoufa and A. Khelil are with Laboratory of Exploitation and Valorization of the Saharan Energetics Resources (LEVRES), Department of Electrical Engineering, Echahid Hamma Lakhdar University, El-Oued, Algeria, email:\{khennoufa-faical, abdellatif-khelil\}@univ-eloued.dz.} 
\thanks{F. Kara is with the Department of Computer Engineering, Zonguldak Bulent Ecevit University, Zonguldak, Turkey, 67100,  e-mail: f.kara@beun.edu.tr.}
\thanks{H. Kaya is with the Electical and Electronics Engineering, Zonguldak Bulent Ecevit University, Zonguldak, Turkey, 67100,  e-mail: hakan.kaya@beun.edu.tr.}
\thanks{X. Li is with School of Physics and Electronic Information Engineering, Henan Polytechnic University, Jiaozuo, China, e-mail: lixingwang@hpu.edu.cn.}
\thanks{K. Rabie is with the Department of Engineering, Manchester Metropolitan University, Manchester M1 5GD, U.K. e-mail: k.rabie@mmu.ac.uk.}
\thanks{H. Yanikomeroglu is with the Department of Systems and Computer Engineering, Carleton University, Ottawa, K1S 5B6, ON, Canada, e-mail: halim@sce.carleton.ca.}

        }
\maketitle
\begin{abstract}
Cooperative non-orthogonal multiple access (CNOMA) has recently been adapted with energy harvesting (EH) to increase energy efficiency and extend the lifetime of energy-constrained wireless networks. This paper proposes a hybrid EH protocol-assisted CNOMA, which is a combination of the two main existing EH protocols (power splitting (PS) and time switching (TS)). The end-to-end bit error rate (BER) expressions of users in the proposed scheme are obtained over Nakagami-$m$ fading channels. The proposed hybrid EH (HEH) protocol is compared with the benchmark schemes (i.e.,  existing EH protocols and no EH). Based on the extensive simulations, we reveal that the analytical results match perfectly with simulations which proves the correctness of the derivations. Numerical results also show that the HEH-CNOMA outperforms the benchmarks significantly. In addition, we discuss the optimum value of EH factors to minimize the error probability in HEH-CNOMA and show that an optimum value can be obtained according to channel parameters. 

\end{abstract}


\IEEEpeerreviewmaketitle
\section{Introduction}
Non-orthogonal multiple access (NOMA) is widely renowned for increasing the number of users and improving spectral efficiency in next-generation wireless networks \cite{Qi2021}. Besides, NOMA can be easily integrated with other physical layer schemes. Hence, the cooperative-NOMA (CNOMA) has been one of the most investigated topics, recently due to its ability to increase spectral efficiency, user fairness and expand network coverage \cite{Vaezi2019a}. On the other hand, to increase the energy efficiency and recharge the battery-limited devices, the energy harvesting (EH) in CNOMA has also received significant attention \cite{Ashraf2021}. In literature, there are two major protocols in EH: power splitting (PS) and time switching (TS). In the PS protocol, the EH and information detection will be performed simultaneously with a PS factor, whereas in TS protocol, the EH and information detection will be implemented at different times slots \cite{Ju2015}.

In order to show the performance gain of EH-NOMA over NOMA without EH, the ergodic capacity and outage probability of simultaneous wireless information and power transfer (SWIPT) aided CNOMA (SWIPT-CNOMA) have been evaluated \cite{Li2019d, Do19, Nayak2021, Do2019, Bisen2021}. Through examining the previous works, although most of the existing EH-CNOMA papers discuss investigating outage probability and ergodic capacity, the error performance (one of the most important key performance indicators (KPIs)) investigation is very limited. In \cite{bariah2019error, mohjazi2016performance,li2019error}, the error performance of CNOMA with EH (with PS protocol) amplify-and-forward (AF) relay has been performed. The authors of \cite{li2020swipt} evaluated the error performance of the $m$th best  relay selection in EH (with PS protocol) AF multiple relay-assisted CNOMA. Then in \cite{Khennoufa2022}, the authors investigate the error performance of CNOMA with a EH relay (with TS protocol). As seen above discussion, the related EH-CNOMA papers deal with only one EH harvesting protocol. However, a comparison of the existing EH protocols is important to select the proper protocol to guarantee the best performance in CNOMA. Regardless superiority of EH to schemes without EH, it is possible to improve EH to reach the best performance. In order to improve the energy efficiency of EH, an optimized protocol called hybrid EH that combines PS and TS jointly has been proposed and evaluated in terms of throughput \cite{Atapattu2016}. According to the findings in \cite{Atapattu2016}, the hybrid protocol outperforms both TS and PS protocols. The hybrid protocol can become one of the important manners to improve EH performance in CNOMA. The hybrid protocol has been evaluated in terms of ergodic capacity and outage probability for CNOMA in \cite{li2020optimal,amin2019performance,amin2021performance}.






 To the best of the authors' knowledge, most of the existing papers investigate the EH-CNOMA schemes by using either TS or PS protocol and no work provides hybrid EH for CNOMA in terms of bit error rate (BER). Nevertheless, the hybrid protocol can improve the performance of EH \cite{Atapattu2016}, and it is considered as a practical scheme, which is a combination of the existing EH protocols. In addition, the BER performance of the EH-CNOMA schemes has not been studied well by considering different EH protocol. Motivated by aforementioned discussions, in this paper, we investigate a hybrid EH protocol in CNOMA (HEH-CNOMA) in terms of BER. The main contributions of the paper are given as follow. \begin{itemize}
    \item We propose HEH-CNOMA scheme where the relay applies a hybrid EH protocol to improve the performance of the existing EH schemes in CNOMA networks.
    \item We also present three benchmark schemes (i.e., PS, TS and no EH) to evaluate the performance of the proposed HEH-CNOMA. We derive the exact end-to-end BER expressions for HEH-CNOMA and three benchmarks over Nakagami-$m$ fading channels. The simulations validate the perfect match of the analysis. Based on the extensive simulations, we reveal that the proposed HEH-CNOMA outperforms all three benchmarks remarkably. 
    \item  We also discuss the optimum values for EH parameters and power allocation coefficient in HEH-CNOMA and we present that optimum values can be obtained to improve performance of both users.
\end{itemize}  

The rest of the paper is presented as follows. The considered HEH-CNOMA system model is introduced along with three benchmark schemes. In Section III we analyze the closed-forms BER of all schemes. The numerical results are presented in Section IV to validate analysis and compare with the benchmarks. Finally, Section V concludes the paper.

\section{System Model}
We consider a CNOMA network consisting of a source (S), a decode-and-forward (DF) relay (R), and two users named user 1 (U$_1$) and user 2 (U$_2$). We assume that the R assists the users to receive the signal since the direct link between source and users does not exist due to large path-loss or obstacles. Each node is equipped with a single antenna. We suppose that the R node works in half-duplex (HD) mode. All communication links are exposed to a Nakagami-$m$ fading. The R node harvests energy from the radio frequency (RF) signal transmitted by the source (i.e., using three EH protocols defined below). In the second phase, the R node forwards the signal to two users using the harvested energy. The transmission block period is $T$. The EH protocols, namely i) PS; ii) TS; and iii) hybrid (i.e., proposed) and their operations are described below.
\begin{itemize}
\item \textbf{PS}: In the first $T/2$ time slot, the received signal's power is split into two parts at the R: The first part ($\rho$) is used to harvest energy, while the second part (1-$\rho$) is utilized to decode data in the decoder (where $\rho$ is the PS factor). The remaining time (i.e., $T/2$) is devoted to delivering the data to the users. 
\item \textbf{TS}: In the first $\beta T$ time slot, the R harvests energy, where $0\leq\beta\leq1$ and $\beta$ is the TS factor. Then, in the first $(1-\beta) T⁄2$ of the remaining time, the S sends a superimposed coding (SC) information to the R. Then, in the last $(1-\beta)T⁄2$ time slot, the R uses the harvested energy to deliver this information signal to the users.
\item \textbf{Hybrid (proposed)}: In the first $\beta T$ time slot, the R harvests energy as being in TS mode, where  $ 0\leq \beta \leq 1$. Then, the S transmits an SC information to the R in the first $(1-\beta)T⁄2$ of the remaining time. Also, during this first $(1-\beta)T⁄2$ time period, the R harvest energy from the received information, where 0 $\leq \rho \leq 1$ ($\rho$ is the  PS factor for energy harvesting during the first $(1-\beta)T⁄2$ at the R). 
Subsequently, the R uses the harvested energy from $\beta T$  plus  $(1-\beta)T⁄2$ time slots to forward the information signal to the users in the last $(1-\beta)T⁄2$.
\item \textbf{No EH}: The R is not capable of harvesting energy, and it uses its own energy source (e.g., battery). The S and the R cover half of the communication (T⁄2 time), and both consume equal power during these data transfers.
\end{itemize}
At the S, the messages of users are combined in an SC information signal with different power levels according to the users' channel gain\footnote{In the NOMA system, the S or R combines the users' messages in an SC signal with different power levels according to the channel gains for each user, where the user with higher channel gain will mostly have less power. Thus, we take into account that the U$_2$ has a higher channel gain than U$_1$, i.e., $|h_{1}|^2<|h_{2}|^2$} and the R receives this information signal in the first phase. Thus, the received information signal by the R node is given as
\begin{equation}\label{eq:1}
y_r= \sqrt{P_{s} \varpi} (\sqrt{\alpha_{1}} s_{1}+\sqrt{\alpha_{2}} s_{2}) h_{r} +n,
\end{equation} \\
where $\varpi$ is the coefficient that indicates how much of the source power is dedicated to information transfer, it will be $(1-\rho)$ for PS and hybrid mode whereas equals to 1 for TS and no-EH modes. $s_{1}$ and $s_{2}$ are the U$_1$ and U$_2$ signals, respectively. $\alpha_1$ and $\alpha_2$ are the power allocation coefficients of U$_1$ and U$_2$, respectively in which $\alpha_1>\alpha_2$, such that $\alpha_1+\alpha_2=1$. $P_{s}$ is the transmit power. $h_{r}$ is the Nakagami-$m$ fading channel coefficient\footnote{The envelope of $h_{r}$ and $h_{k}$ follows Nakagami-$m$ fading with $\Omega_r$ and $\Omega_k$ spread and $m_{r}$ and $m_{k}$ shape parameters.} between S-R. $ n \sim \mathcal{CN} (0,\sigma^2)$ is the additive white Gaussian noise (AWGN) at R. When the information signal is received, the $R$ node decodes the $s_{1}$ firstly. Then, through the SIC, $s_{2}$ is recovered.

For fairness, we consider the total consumed energy in all protocols is equal to $\epsilon_{T}$. Thus, $\epsilon_{T}$ is consumed equally by the S and R in the no-EH mode, while it is consumed only by the S in EH protocols, the source power is expressed as
\begin{equation}\label{eq:2}
   P_s=\phi P_T, \ \text{where}\ P_T=\epsilon_{T}/T,
\end{equation}
where $\phi$ is equal to 1 for no-EH, 2 for PS and $\frac{2}{\beta+1}$ for TS and hybrid. 
In EH protocols, the harvested energy during the power transfer in each protocol  is given as
\begin{equation}\label{eq:3}
   E_{h}=\left \{ \begin{array}{lll}
   \eta P_{s} \rho \frac{T}{2}|h_{r}|^2,  & \mbox{PS},  \\
   \eta P_{s} \beta T |h_{r}|^2,  & \mbox{TS},\\
   \eta P_{s} (\frac{2 \beta + \rho (1-\beta)T)}{2}) |h_{r}|^2,  & \mbox{Hybrid},\\
   \end{array} \right. 
\end{equation}
where $\eta$ ($0 <\eta<1$) is the energy conversion efficiency factor. 
Thus, the transmit power of the relay through the harvested energy is given by
  \begin{equation}\label{eq:4}
   P_{r}=\left \{ \begin{array}{ll}
   P_{s} |h_{r}|^2 \Psi,  & \mbox{PS, TS and Hybrid},\\
   P_{s},  & \mbox{no EH}, 
   \end{array} \right. 
\end{equation}
where
\begin{equation}\label{eq:5}
   \Psi=\left \{ \begin{array}{lll}
   \eta \rho,  & \mbox{PS},\\
   \sfrac{2 \eta \beta}{(1-\beta)},  & \mbox{TS}, \\
   \eta (\rho+\sfrac{2 \beta}{(1-\beta)}),  & \mbox{Hybrid}. 
   \end{array} \right. 
\end{equation} 

Based on the received information signal, the R implements maximum likelihood detection (MLD)/successive interference cancellation (SIC), to detect $s_{1}$ and $s_{2}$ symbols respectively.Based on the signals decoded in the previous phase, the R implements a new SC signal and by using the harvested energy, it forwards this new SC signal to the users. The received information at the users is given by
\begin{equation}\label{eq:6}
y_k= \sqrt{P_r\varpi} (\sqrt{\alpha_{1}} s_{1}+\sqrt{\alpha_{2}} s_{2}) h_{k} +n ,\  k={
1, 2},
\end{equation} 
where $h_{k}$, $k={1,2}$ is the Nakagami-$m$ fading channel coefficient between $R-$U$_{k}$. Through the information signal received by users, the U$_1$ detects its symbols $s_{1}$ directly, while the U$_2$ use the SIC to detect its symbols $s_{2}$.

\section{Performance analysis}
In this section, over Nakagami-$m$ fading channels, we derive the closed-form BER of the two phases. Then, we obtain the end-to-end (e2e) BER expressions for each user.
At both S and R, a binary phase-shift keying (BPSK) modulation is utilized to obtain the $s_{1}$ and $s_{2}$ signals. Hence, two different energy levels with similar probabilities could be presented. By using a MLD, for all EH protocols and no EH, the R detects $s_{1}$ firstly by treating $s_{2}$ as noise. Thereafter, $s_{2}$ is detected using the SIC. Then, the R implements an SC signal again for detected $s_{1}$ and $s_{2}$ symbols and forwards it to users by using the harvested energy (i.e., for the case of the EH protocols) or its own power (i.e., case of no EH). Through MLD, U$_1$ detects its own symbols $s_{1}$, whereas U$_2$ uses a SIC to identify its symbols $s_{2}$. By using the common probability of independent error event in two phases, the e2e BER at each user is given as in \cite{babaei2018ber} by
\begin{equation}\label{eq:7}
P_{e2e,k} (e)=1-\left.(1-P_{r, k} (e)\right.) \left.(1-P_{k} (e)\right.),\ k={1,2},
\end{equation} 
where $P_{r, k} (e)$ is the BER of U$_{k}$ at the relay. $P_{k} (e)$ is the BER at U$_{k}$ in the second phase. 
 \subsection{BER in the first phase}
Based on the discussion above (i.e., for EH and no EH), R detects $s_{1}$ in the first step. The BER of $s_{1}$ at R, by following the steps in \cite{khennoufa2022bit}, can be given as
\begin{equation}\label{eq:8}
P_{r,1} (e)=\frac{1}{2} \sum_{{i}=1}^{2} \mathrm{Q} ( \sqrt{\varpi \gamma \zeta_{i} |h_{r}|^2}), 
\end{equation} 
where $\gamma=\frac{P_{s}}{\sigma^2}$, $\zeta_{i}=[(\sqrt{\alpha_1}+\sqrt{\alpha_2})^2,(\sqrt{\alpha_1}-\sqrt{\alpha_2})^2]$.
The average BER (ABER) of (8) over a Nakagami-$m$ fading channel by using \cite{Simon2005} is given as
\begin{equation}\label{eq:9}
\begin{split}
 & P_{r,1} (e)=\frac{1}{2} \sum_{{i}=1}^{2} \frac{1}{\sqrt{2 \pi}} \frac{\Gamma({m_{r}+0.5})}{\Gamma(m_{r}+1)}  \\ &   \times (\frac{\sqrt{\varpi \gamma \zeta_{i} \delta /2 m_{r}}}{(1+\sqrt{\varpi \gamma \zeta_{i} \delta /2 m_{r}})^{(m_{r}+0.5)}})  \nabla_1(m_{r},\varpi \gamma \zeta_{i} \delta),\\
 \end{split}
 \end{equation}
where $\nabla_1(m_{r},\varpi \gamma \zeta_{i} \delta)=_2F_1(1,m_{r}+0.5,m_{r}+1,\frac{2 m_{r}}{2 m_{r}+\varpi \gamma \zeta_{i} \delta})$ and $\delta=\Omega_{r}$. $_2F_1(.,.;.,.)$ is the Gauss hypergeometric function.

We also need to analyze the BER performance of the correct and erroneous SIC as \cite{khennoufa2022bit} to find the BER of $s_{2}$ at R. Thus, the BER performance of $s_{2}$ at R is given by
 
\begin{equation}\label{eq:10}
P_{r,2} (e)=\frac{1}{2} \sum_{{j}=1}^{5} \nu^{j} \mathrm{Q} (\sqrt{\varpi \gamma \zeta_{j} |h_{r}|^2}), 
\end{equation} 
where $\nu^{j}=[2,-1,1,1,-1]$, $\zeta_{j}=[\alpha_2, (\sqrt{\alpha_1}+\sqrt{\alpha_2})^2, (\sqrt{2 \alpha_1}+\sqrt{\alpha_2})^2,(\sqrt{\alpha_1}-\sqrt{\alpha_2})^2, (2 \sqrt{\alpha_1}-\sqrt{\alpha_2})^2]$.
By using \cite{Simon2005}, the ABER of (10) over the Nakagami-$m$ fading channel can be given as
\begin{equation}\label{eq:11}
\begin{split}
 & P_{r,2} (e)=\frac{1}{2} \sum_{{j}=1}^{5} \nu^{j} \frac{1}{\sqrt{2 \pi}} \frac{\Gamma({m_{r}+0.5})}{\Gamma(m_{r}+1)}   \times \\ & (\frac{\sqrt{\varpi \gamma \zeta_{j} \delta /2 m_{r}}}{(1+\sqrt{\varpi \gamma \zeta_{j} \delta /2 m_{r}})^{(m_{r}+0.5)}})  \nabla_2(m_{r},\varpi \gamma \zeta_{j} \delta), \\
  \end{split}
 \end{equation}
 where $\nabla_2 (m_{r},\varpi \gamma \zeta_{j} \delta)=_2F_1(1,m_{r}+0.5,m_{r}+1,\frac{2 m_{r}}{2 m_{r}+\varpi \gamma \zeta_{j} \delta})$.
\subsection{BER in the second phase}
Using the harvested energy (i.e., using the power harvested of the three EH protocols as presented in section II), the R node forwards the SC signal to the users. The U$_1$ detect its symbols directly using MLD, so the BER at U$_1$ can be given by
\begin{equation}\label{eq:12}
P_{1} (e)=\frac{1}{2} \sum_{{i}=1}^{2} \mathrm{Q} (\sqrt{\gamma \zeta_{i} \psi |h_{1}|^2|h_{r}|^2}). 
\end{equation} 
As seen in (12), the conditional BER includes multiplying of two random variables which belong to first and second phase link qualities. Although, the conditional BER in (12) defines the BER of second phase, it includes $|h_r|^2$ since the transmit power of the relay (i.e., $P_r$) is a function of it (4). Thus, the ABER of (12) can not be simply obtained by averaging Gamma distribution (i.e., the absolute square of Nakagami-$m$). Firstly, the joint probability density function (PDF) of $|h_1|^2|h_r|^2$ should be derived. By following steps in \cite{Babaei2018} through the PDF of the product of two Chi-Square distributions, the PDF can be obtained.  After simplification and using \cite[eq. (3.478.4)]{Gradshteyn1994}, we derive the ABER of (12) by using the Meijer-G function \cite[eq.(9.311)]{Gradshteyn1994} by following the steps of \cite{Babaei2018,Kara21}, and  the closed-form ABER at U$_1$ is given as 
\begin{equation}\label{eq:13}
P_{1} (e)=\frac{1}{2} \sum_{{i}=1}^{2} \frac{\Delta_1}{2 \sqrt{\pi}} \MeijerG{3}{3}{4}{5}{0, 1-u_{2}, 0.5-u_{2},1-u_{2}}{0.5 u_{1},-0.5 u_{1},1-u_{2}, u_{2},0 }{\frac{\Delta_{3}}{4}},
\end{equation} 
where $u_{1}=m_{r}-m_{1}, u_{2}=0.5(m_{r}+m_{1}), \Delta_{1}=\Delta_{2} (\frac{m_{1} \Omega_{r}}{m_{r} \Omega_{1}})^{\frac{u_{1}}{2}} (\frac{1}{\gamma \zeta_{i} \psi})^{u_{2}}, \Delta_{2}= \frac{(\frac{m{r}}{\Omega_{r}})^{m{r}} (\frac{m{1}}{\Omega_{1}})^{m{1}}}{\Gamma (m_{r}) \Gamma (m_{1})}, \Delta_{3}=2 \sqrt{\frac{m_{1} m_{r}}{\Omega_{1} \Omega_{r} \gamma \zeta_{i} \psi}}$, $m_{r}$ and $m_{k}$ are the Nakagami fading parameters.

Also, we obtain the BER at U$_2$ through the correct and erroneous SIC. Thus, the BER at U$_2$ is given by
\begin{equation}\label{eq:14}
P_{2} (e)=\frac{1}{2} \sum_{{j}=1}^{5} \nu^{j} \mathrm{Q} (\sqrt{\gamma \zeta_{j} \psi |h_{2}|^2|h_{r}|^2}). 
\end{equation} 
As discussed above, the ABER of (14) needs further modification. Thus, we find the ABER of (14) by using the same method of (13) and it is given as
\begin{equation}\label{eq:15}
P_{2} (e)=\frac{1}{2} \sum_{{j}=1}^{5} \nu^{j} \frac{\Delta_4}{2 \sqrt{\pi}} \MeijerG{3}{3}{4}{5}{0, 1-u_{4}, 0.5-u_{4},1-u_{4}}{0.5 u_{3},-0.5 u_{3},1-u_{4}, u_{4},0 }{\frac{\Delta_{6}}{4}},
\end{equation} 
where $u_{3}=m_{r}-m_{2}, u_{4}=0.5(m_{r}+m_{2}), \Delta_{4}=\Delta_{5} (\frac{m_{2} \Omega_{r}}{m_{r} \Omega_{2}})^{\frac{u_{1}}{2}} (\frac{1}{\gamma \zeta_{j} \psi})^{u_{4}}, \Delta_{5}= \frac{(\frac{m{r}}{\Omega_{r}})^{m{r}} (\frac{m{2}}{\Omega_{2}})^{m_{2}}}{\Gamma (m_{r}) \Gamma (m_{2})}$ and $ \Delta_{6}=2 \sqrt{\frac{m_{2} m_{r}}{\Omega_{2} \Omega_{r} \gamma \zeta_{j} \psi}}$.

Finally, we obtain the e2e ABER expression of U$_1$ as (16) by substituting (9) and (13) into (7). Also, by substituting (11) and (15) into (7), we get the e2e ABER expression of U$_2$ as (17) (please see the top of the next page). 

\begin{figure*}
   \begin{equation}\label{eq:16}
   \begin{split}
& P_{e2e,1}(e)= \\
 &1-\left(1-\frac{1}{2} \sum_{{i}=1}^{2} \frac{1}{\sqrt{2 \pi}} \frac{\Gamma({m_{r}+0.5})}{\Gamma(m_{r}+1)}\times \frac{\sqrt{\varpi \gamma \zeta_{i} \delta /2 m_{r}}}{(1+\sqrt{\varpi \gamma \zeta_{i} \delta /2 m_{r}})^{(m_{r}+0.5)}} \nabla_1(m_{r},\varpi \gamma \zeta_{i} \delta)\right) \left(1-\frac{1}{2} \sum_{{i}=1}^{2} \frac{\Delta_1}{2 \sqrt{\pi}} \MeijerG{3}{3}{4}{5}{0, 1-u_{2}, 0.5-u_{2},1-u_{2}}{0.5 u_{1},-0.5 u_{1},1-u_{2}, u_{2},0 }{\frac{\Delta_{3}}{4}}\right)\\
 \end{split}
\end{equation}
\hrulefill
\end{figure*}

\begin{figure*}
 \begin{equation}\label{eq:17}
\begin{split}
 & P_{e2e,2} (e)= \\
 &1-\left(1-\frac{1}{2} \sum_{{j}=1}^{5} \nu^{j} \frac{1}{\sqrt{2 \pi}} \frac{\Gamma({m_{r}+0.5})}{\Gamma(m_{r}+1)} \times \frac{\sqrt{\varpi \gamma \zeta_{j} \delta /2 m_{r}}}{(1+\sqrt{\varpi \gamma \zeta_{j} \delta /2 m_{r}})^{(m_{r}+0.5)}}\nabla_2(m_{r},\varpi \gamma \zeta_{j} \delta)\right) \left(1-\frac{1}{2} \sum_{{j}=1}^{5} \nu^{j} \frac{\Delta_4}{2 \sqrt{\pi}} \MeijerG{3}{3}{4}{5}{0, 1-u_{4}, 0.5-u_{4},1-u_{4}}{0.5 u_{3},-0.5 u_{3},1-u_{4}, u_{4},0 }{\frac{\Delta_{6}}{4}}\right)\\
\end{split}
\end{equation}
\hrulefill
\end{figure*}

 \section{Numerical Results}
In this section, we validate the theoretical analysis by Monte Carlo simulation. In all figures, the lines represent the simulation results, while the markers represent the analytical results. The channel parameters for all simulations are given in Table I and in all simulations, $\eta=0.95$ and for a fair comparison, total transmit SNR is defined as $\sfrac{P_T}{N_0}$. In simulations, we use normalized {i.e., T=1} transmission time.

In Fig. 1, we present ABER performances of both users for all protocols, where $\alpha_1=0.9$, and $\beta=\rho=0.1$. Firstly, it is worthy noting that our analysis matches perfectly with simulations in all protocols and scenarios which proves the correctness of our closed-from expressions. Besides, the analysis refers to both integer and non-integer values of $m$ and it can be seen that this shape parameter $m$ drives the diversity order. In Fig. 1, we can observe that the proposed HEH-CNOMA outperforms all three benchmarks. By examining the comparisons in Fig. 1, we observe that the performance in Scenario I is better than Scenario III and in the same manner, the performance in Scenario IV is superior to  Scenario II. These observations are a natural expectation of channel quality differences in various scenarios. The shape parameter in Scenario I is greater than Scenario III so that Scenario I has  better ABER performance and improved diversity order since in both phases less erroneous detection is performed. Likewise, Scenario IV outperforms Scenario II since the spread  parameters are higher which improves the performance. On the other hand, please note that in Scenario I and Scenario IV, the linear sum of spread parameters (i.e., $\Omega_r+\Omega_k$) of links are fixed which can reflect the relay position. In other words, the higher $\Omega_r$ means that the relay is closer to the source than the users. By comparing the performances of Scenario I and Scenario IV, we can observe that Scenario IV is superior. Thus, we can easily say that the CNOMA schemes have better performance when the relay is close to the middle of the source and users. 
\begin{table}[]
\centering
\caption{Channel parameters in simulations}
\begin{tabular}{|c|c|c|c|c|c|}
\hline
&Scenario I&Scenario II&Scenario III&Scenario IV \\ \hline
$\left[\Omega_r, \ \Omega_1,\ \Omega_2\right]$& $[10,\ 2,\ 10]$& $[2,\ 2,\ 10]$ &$[10,\ 2,\ 10]$&$[8,\ 4,\ 12]$\\ \hline
$m_r=m_1=m_2$& $1.5$& $1.5$ &$1$&$1.5$\\ \hline
\end{tabular}
\end{table}
\begin{figure}
    \centering
    \subfloat[{Comparison between Scenarios I and III}]{\includegraphics[width=.8\columnwidth]{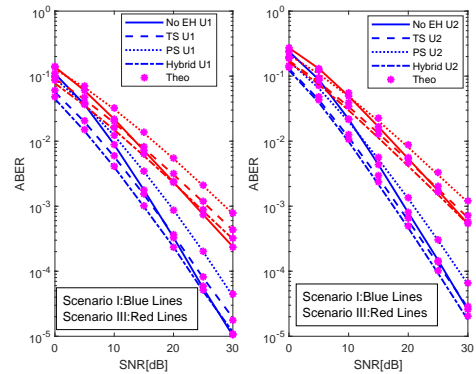}
  \label{img1:const1}}\\
 \centering
    \subfloat[{Comparison between Scenarios II and IV}]{\includegraphics[width=.8\columnwidth]{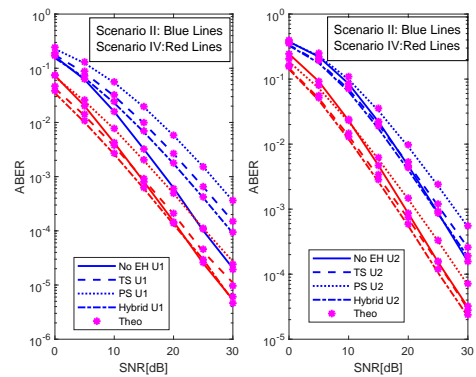}
\label{img1:const1}}
    \caption{ABER performance comparisons between HEH-CNOMA and benchmarks w.r.t. SNR.}
    \label{constellations}
\end{figure}


\begin{figure}
    \centering
    \includegraphics[width=0.8\columnwidth]{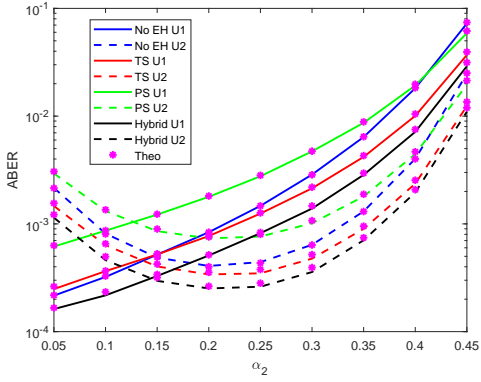}
    \caption{ABER performance comparisons between HEH-CNOMA and benchmarks w.r.t. power allocation.}
    \label{constellations}
\end{figure}

\begin{figure}
    \centering
\subfloat[{Scenario I}]{\includegraphics[width=.78\columnwidth]{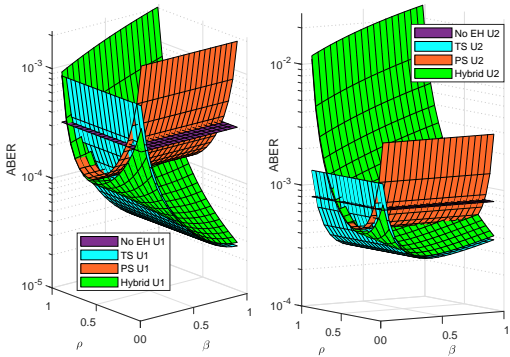}
\label{img1:const1}}\\
\centering
\subfloat[{Scenario IV}]{\includegraphics[width=.78\columnwidth]{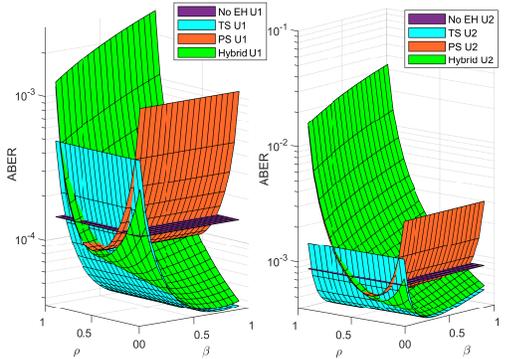}
\label{img1:const2}}
    \caption{ABER performance comparisons between HEH-CNOMA and benchmarks w.r.t. EH parameters.}
    \label{constellations}
\end{figure}


The power allocation has prevailing effects on the ABER performance. To this end, to discuss the optimal power allocation (PA), in Fig. 2, we present ABER performances of users in Scenario I for $\sfrac{P_T}{N_0}=20$ dB. Regardless of the PA changes, the HEH-CNOMA still outperforms all three benchmarks. Nevertheless, as expected, with the increase of $\alpha_2$, the performance of U$_1$ is always decayed (i.e., ABER is increasing) since $\alpha_1$ so the allocated power to U$_1$ symbols is reduced. On the other hand, by increasing $\alpha_2$, we firstly have better performance for `U$_2$; however, it becomes worse with so much increase in $\alpha_2$. This can be explained as follows. By increasing $\alpha_2$, we allocate more power to U$_2$ symbols so that the performance is improved. However, U$_2$ should implement SIC (i.e., detect U$_1$ symbols firstly and subtract it) to detect its symbols. Therefore, if we increase $\alpha_2$, U$_1$ symbols will be detected erroneously during SIC at the U$_2$ and it will cause an error floor in detection of U$_2$ symbols. Based on the above discussion and Fig. 2, by considering both users' performances, we can say that the optimum value of power allocation should be $0.1\leq\alpha_2\leq0.25$.

In order to better illustrate the effects of the EH parameters ($\beta$, $\rho$) on the performance, in Fig. 3, we present the ABER versus $\beta$ and $\rho$ in Scenario I and Scenario IV. In both comparisons, we can say that an optimal $\beta$-$\rho$ pair can be obtained to minimize the ABER in HEH-CNOMA. According to channel conditions, $\beta$ and $\rho$ can be chosen jointly. For instance, in Scenario I, since the relay is close to the source, the link between $S$-$R$ has already a good quality and less erroneous detection is performed at the relay. Hence, we can increase $\beta$ and $\rho$ to increase the harvested energy to guarantee a reliable communication in the second phase by increasing the transmit power of the relay. On the other hand, when the relay is located nearby the middle (i.e., Scenario IV), we should select lower $\beta$ and $\rho$ values. If we choose $\beta$ and $\rho$ too much, the information processing in the first phase becomes unreliable since less power is devoted to information transfer. In addition to this, the channel quality of the first phase is relatively worse, so the relay makes more error in detection and this causes an error propagation from relay to users.


\section{Conclusion}
This paper proposes a hybrid protocol-assisted CNOMA, which is a combination of existing EH protocols (i.e., PS and TS). The aim of the study is to improve the performance of the existing EH protocols with a more realistic protocol for practical assumptions. The e2e BER expressions of users are analyzed over the Nakagami-$m$ fading channel. We discuss the optimal hybrid protocol values with the power allocation factor. The results illustrate that proposed HEH-CNOMA outperforms all benchmarks. The optimum value of EH factors to achieve more satisfactory performance for both users jointly in HEH-CNOMA can be obtained according to spread and shape parameters of the channel.

\bibliographystyle{IEEEtran}
\bibliography{references}
\end{document}